\documentclass[aps,twocolumn,prb,superscriptaddress,showpacs]{revtex4}
\usepackage{amsfonts}
\usepackage{amsmath}
\usepackage{amssymb}
\usepackage{graphicx}
\usepackage{color}
\newcommand{\bq}{\begin{equation}}
\newcommand{\eq}{\end{equation}}
\begin{document}
\preprint{}
\title{Electron spin resonance in Kondo systems}
\author{Elihu Abrahams}
\affiliation{Center for Materials Theory, Serin Physics Laboratory, Rutgers University,
Piscataway, NJ 08854-8019}
\author{Peter W{\" o}lfle}
\affiliation{Institut f{\" u}r Theorie der Kondensierten Materie, Universit{\" a}t
Karlsruhe, D-76128 Karlsruhe, Germany}

\date{\today}

\begin{abstract}
We calculate the dynamical spin response of Kondo impurity and Kondo lattice
systems within a semiphenomenological Fermi liquid description, at low
temperatures $T<T_K$, the Kondo temperature, and low magnetic fields $B \ll
k_B T_K/g\mu_B$. Fermi liquid parameters are determined by comparison (i)
with microscopic theory (numerical renormalization group) for the impurity
model and (ii) with experiment for
the lattice model. We find in the impurity case that the true impurity spin
resonance has a width of the order of $T_K$ and disappears altogether if the 
$g$-factors of impurity spin and conduction electron spin are equal.
However, there is an impurity-induced resonance contribution at the
conduction electron resonance. The latter is broadened by spin lattice
relaxation and is usually unobservable. In contrast, for the Anderson
lattice in the Kondo regime we find a \textit{sharp} ESR resonance line only
slightly shifted from the local resonance and broadened by spin lattice
relaxation, the latter significantly reduced by both the effects of heavy
fermion physics and ferromagnetic fluctuations. We conjecture that our
findings explain the sharp ESR-lines recently observed in several heavy
fermion compounds.
\end{abstract}

\maketitle


\section{Introduction}

The Kondo effect is arguably the best-studied many-body effect in condensed
matter physics \cite{hewson}. In its initial form \cite{Kondo,Wilson}, it
involves a local ``impurity" spin in a $d$- or $f$-orbital,
antiferromagnetically coupled to the spins of a conduction band in a dilute
magnetic alloy. At temperatures $T$ below the dynamically generated energy
scale $T_{K}$, the Kondo temperature, this interaction causes a local spin $%
1/2$ to be fully screened. This behavior should be noticeable in the $T$%
-dependence of the spin dynamics of the system, as probed by electron spin
resonance (ESR). In fact, the local spin resonance in dilute Kondo compounds
at $T\gg T_{K}$ had been observed even before the Kondo effect was
understood. Afterward there were a number of systematic experimental
investigations and perturbative calculations for the ESR at $T\gg T_{K}$ in
dilute Kondo systems \cite{sweer}.

At low temperatures $T\ll T_{K}$, on the other hand, neutron scattering
studies revealed the existence of a broad spin excitation peak of width $%
T_{K}$, interpreted as the Kondo bound state \cite{neutron}. Within the
isotropic $s$-$d$ exchange (Kondo) model the total spin is conserved.
Therefore, in the limit of equal $g$-factors of local moments and conduction
electron spins one expects a single spin resonance line at all temperatures,
only broadened by spin lattice relaxation. As we shall show below, in this
limit the weight of the broad local spin resonance tends to zero.

In several recent experiments \cite{YRS1,Ce} low-temperature ESR has been
observed in some heavy-fermion metals, in particular YRh$_2$Si$_2$ (YRS) 
\cite{YRS2}. The phase diagram of YRS has a magnetic-field induced
quantum critical point and is a model system for the study of quantum
criticality in the Kondo lattice. Consequently, the observation of a narrow
ESR resonance in this compound aroused great interest, especially since it
was commonly believed that heavy-fermion ESR would be unobservable due to an
enormous intrinsic linewidth $\Delta B$ of order $k_BT_K/g\mu_B$ \cite{YRS1}%
. Here $T_K$ is the lattice coherence (``Kondo") temperature for the onset
of heavy-fermion behavior and $g\mu_B$ is the gyromagnetic ratio for the
resonance. These were the first observations of ESR in Kondo lattice systems
at $T < T_K.$

In YRS, the observed narrow dysonian \cite{feher} ESR line shape was
originally interpreted \cite{YRS1} as indicating that the resonance was due
to local spins at the Yb sites. Therefore, initially the authors speculated
that the appearence of a narrow ESR line might indicate the suppression of
the Kondo effect near the quantum critical point, since, as explained above,
carrying over Kondo impurity physics to the Kondo lattice one might expect
the local spins to be screened by the Kondo effect, giving rise only to a
broad spin excitation peak, too wide to be observed in ESR experiments.
However, a closer look \cite{ea} revealed that itinerant (heavy) electron
ESR could give rise to a similar line shape since the carrier diffusion in
YRS is quite slow. Thus, whether the resonance was that of localized or
itinerant spins remains an open question.

Now, a common feature of the compounds in which ESR has been observed
appears to be the existence of ferromagnetic fluctuations \cite{Ce} These
findings challenge our understanding of heavy fermion compounds: How does a
sharp electron spin resonance emerge despite Kondo screening and spin
lattice relaxation, and why is this process influenced by ferromagnetic
fluctuations? We shall address these questions in the framework of Fermi
liquid theory, taking the relevant parameters from numerical studies and
experiment.

\section{Anderson impurity model in the Kondo screened regime}

In the Kondo regime an impurity spin is screened by the conduction electron
spins at (or near) the impurity. The dynamics of the impurity spin is
governed by the energy scale of the corresponding many-body resonance, the
Kondo temperature $T_{K}$. Nonetheless the conduction electrons in the
vicinity of the impurity show the influence of the Kondo screened state in
their dynamical behavior. In the Anderson model, the local spin is that of a
localized $f$ electron
We assume that the Zeeman splittings $\omega_{f}$
and $\omega_{c}$ induced by a magnetic field acting on the local and
conduction electron spins are small compared to the Kondo temperature $T_{K}$%
. Then the Kondo screened state is only weakly perturbed by the magnetic
field. At temperatures $T\ll T_{K}$, the spin resonance behavior of the
impurity may then be described by Fermi liquid theory \cite{pn}.

We start from the bare Anderson model Hamiltonian 
\begin{eqnarray}
H=H_{c}&+&\sum_{\mathbf{k},\sigma} \epsilon_{\mathbf{k}\sigma}c_{\mathbf{k}%
\sigma}^{+}c_{\mathbf{k}\sigma}+\sum_{\sigma}
\epsilon_{f\sigma}n_{f\sigma}+Un_{f\uparrow}n_{f\downarrow} \nonumber \\
&+& V \sum_{\mathbf{%
k},\sigma} (f_{\sigma}^{+}c_{\mathbf{k}\sigma}+h.c.),
\end{eqnarray}
where $H_{c}$ is the conduction electron Hamiltonian and $c_{\mathbf{k}%
\sigma}^{+},f_{\sigma}^{+}$ are creation operators of the conduction
electrons in momentum and spin eigenstates $(\mathbf{k}\sigma)$, and of
electrons in the local $f$ level, respectively. The operator $%
n_{f\sigma}=f_{\sigma}^{+} f_{\sigma}$ counts the number of electrons on the
local level, and\ $\epsilon _{f\sigma}=\epsilon_{f}-\omega_{f}\sigma/2,$ $%
\sigma=\pm1$.

The effect of the interaction $U$ is to renormalize the parameters $\epsilon
_{f\sigma },U,V$ \ to $\widetilde{\epsilon }_{f\sigma },\widetilde{U},%
\widetilde{V}$ in the renormalized Fermi liquid type low energy Hamiltonian,
Eq. (1) with the renormalized parameters that may be calculated using the
numerical renormaliztion group (NRG) method \cite{NRG}. To keep the algebra
simple, we assume particle-hole symmetry in the following. Then $\widetilde{%
\epsilon }_{f\sigma }=-(\widetilde{U}+\omega _{f}\sigma )/2$. The
hybridization of the local level with the conduction band leads to an $f$%
-level broadening $\widetilde{\Gamma }=\pi \widetilde{V}^{2}N_{0}\sim T_{K}$
with $N_{0}=1/W$ the local conduction electron density of states (DOS) at
the Fermi level (in the model with flat DOS, $W$ is the bandwidth). The
initially rather large bare level width is renormalized down to the very
narrow width of the Kondo resonance. The NRG calculation shows that $%
\widetilde{U}=\pi \widetilde{\Gamma }$ and $\widetilde{V}^{2}$ is O$%
(T_{K}/T_{F})$, where $T_{F}$ is the Fermi temperature of the conduction
electrons.

In the framework of Fermi liquid theory, the interaction has two major
consequences: (i) it gives rise to a molecular field renormalizing the
collective response of the system (ii) it leads to a finite lifetime of
quasiparticles. However, the quasiparticle relaxation rate is limited by the
available phase space and vanishes quadratically as the excitation energy
goes to zero. Therefore, at temperatures $T\ll T_{K}$ the Landau
quasiparticles are well-defined. The quasiparticle decay contributes to the
spin relaxation rate. As we shall show, the local moment relaxation is
governed by rapid spin flips on the frequency scale of $T_{K}$, occuring as
part of the many body resonance. Then at temperatures $T\ll T_{K}$ we may
neglect the additional relaxation caused by the quasiparticle decay.

We now consider the effects of the molecular field caused by the Fermi
liquid interaction $\widetilde{U}$. We treat the interaction term in the
Hamiltonian in mean field approximation: $\widetilde{U}n_{f\uparrow
}n_{f\downarrow }\approx \frac{1}{2}\widetilde{U}[\langle n_{f}\rangle
n_{f}-\langle m_{f}\rangle m_{f}+const]$, where we defined the density and
spin density operators $n_{f}=n_{f\uparrow }+n_{f\downarrow }$ and \ $%
m_{f}=n_{f\uparrow }-n_{f\downarrow }$. In the case of particle-hole
symmetry, when $\langle n_{f}\rangle =1$, the density term is cancelled by
the single-particle energy. The spin-density term gives rise to an effective
magnetic field so that $\widetilde{\epsilon }_{f\sigma }+\frac{1}{2}%
\widetilde{U}[\langle n_{f}\rangle -\sigma \langle m_{f}\rangle ]=-\sigma
\omega _{f}$, which amounts to a doubling of the Zeeman energy. Here we have
used that the spin polarization is given by \ $m=\langle m_{f}\rangle = \chi
_{ff}^{+-}(0)\omega _{f}/2=\omega _{f}/\pi \widetilde{\Gamma }$, with $\chi
_{ff}^{+-}(0)=2/\pi \widetilde{\Gamma }$ the static susceptibility of the
local spin and $\widetilde{U}=\pi \widetilde{\Gamma }$, as obtained from
NRG calculations. Then the local electron Green's function, including the
coupling to the conduction electrons, is given by 
\begin{equation}
G_{f\sigma }(i\omega _{n})=[i\omega _{n}+\sigma \omega _{f}+i\widetilde{%
\Gamma }sign(\omega _{n})]^{-1}
\end{equation}%
The local spectral function is given by $A_{f\sigma }(\omega )=\mathrm{Im}\
G_{f\sigma }(\omega +i0)=\widetilde{\Gamma }/[(\omega +\sigma \omega
_{f})^{2}+\widetilde{\Gamma }^{2}]$, describing the Kondo resonance. We see
that in a magnetic field the resonance is shifted from its zero field
position $\omega =0$ to the spin dependent position $\omega =-\sigma \omega
_{f}$, which is double the Zeeman shift.

We use the definitions $\omega _{f}=g_{f}\mu _{B}B$, $\omega _{c}=g_{c}\mu
_{B}B$ and take $\chi _{ff}^{+-}$, etc. to be response functions of spin $1/2
$ operators. The dynamical conduction electron susceptibility $\chi _{cc}$
is characterized by a resonance peak broadened by spin-lattice relaxation.
We follow Barnes and Zitkova-Wilcox \cite{Barnes} and model the spin-lattice
relaxation mechanism by a local random magnetic field $\mathbf{{h_{i}}}$
that fluctuates in both direction and magnitude. Then the conduction
electron Hamiltonian is 
\begin{eqnarray}
H_{c}&=&\sum_{\mathbf{k},\sigma }\epsilon _{\mathbf{k}\sigma }c_{\mathbf{k}%
\sigma }^{+}c_{\mathbf{k}\sigma } \nonumber \\
&+&\sum_{\mathbf{k},\sigma ,\mathbf{k\prime
,\sigma \prime }}\sum_{i}\mathbf{h}_{i}\cdot c_{\mathbf{k}\sigma }^{+}%
\boldsymbol{\sigma }_{\sigma \sigma \prime }c_{\mathbf{k\prime }\sigma
\prime }\ e^{i(\mathbf{k}-\mathbf{k^{\prime }})\cdot \mathbf{R}_{i}}
\end{eqnarray}%
The random field components are assumed to be Gaussian correlated, with $%
\langle \mathbf{h}_{i}\rangle =0$ and $\langle h_{i}^{m}h_{j}^{n}\rangle
=\delta _{ij}\delta _{mn}h^{2}$. In Born approximation the average
conduction electron Green's function is then given by 
\begin{equation}
G_{c\uparrow }^{0}(k,i\omega _{n})=[i\omega _{n}-\varepsilon _{k}+\omega
_{c}/2+i\gamma \mathrm{sign}(\omega _{n})]^{-1},
\end{equation}%
where $\gamma =\pi N_{0}h^{2}$.

The impurity induced component of the dynamical transverse susceptibility $%
\chi _{imp}^{+-}(\Omega )$, where $\Omega $ is the frequency of an a.c.
electromagnetic field polarized transverse to the static magnetic field, is
a sum of three contributions, from the conduction electrons (cc), the local
electrons (ff) and the mixed response of conduction electrons to a spin
polarization of the local electrons or vice versa (cf): 
\begin{eqnarray}
\chi _{imp}^{+-}(\Omega )&=&\mu _{B}^{2}\{g_{f}^{2}\chi _{ff}^{+-}(\Omega
)+2g_{c}g_{f}\chi _{cf}^{+-}(\Omega) \nonumber \\
&+& g_{c}^{2}[\chi _{cc}^{+-}(\Omega
)-\chi _{cc}^{bulk}(\Omega )]\}
\end{eqnarray}%
The partial susceptibilities may be calculated using standard many body
techniques, see the Appendix. One finds resonances at the two frequencies $\omega _{f}$ and $%
\omega _{c}$. They have very different widths: the local electron spin
resonance is broadened by $\widetilde{\Gamma }$, whereas the bulk and the
impurity induced conduction electron resonances are broadened by $4\gamma$.
The results are given in the Appendix, Eqs.\ (A1-A3).
Assuming that $\widetilde{\Gamma }>>4\gamma $, it makes sense to consider
the behavior at higher frequencies $\Omega \gg (\omega _{f,c},\gamma )$
(regime I) and low frequencies (regime II) separately. In regime I one
finds: 
\begin{equation}
\chi _{imp}^{+-}(\Omega )=\frac{2\mu _{B}^{2}(g_{f}-g_{c})^{2}}{\pi 
\widetilde{\Gamma }}\frac{i\widetilde{\Gamma }}{\Omega -\omega _{f}+i%
\widetilde{\Gamma }}.
\end{equation}%
Neutron scattering data on diluted magnetic alloys show a broad resonance in 
$\chi _{imp}^{+-}(\Omega )$ of width $T_{K}$ \cite{neutron}, in accordance
with the above result. Note that this broad peak vanishes in the case of
equal g-factors, as a consequence of the conservation of magnetization in
that case (leaving aside spin lattice relaxation for the moment). The result
in regime II is: 
\begin{eqnarray}
\chi _{imp}^{+-}(\Omega )&=&\frac{2\mu _{B}^{2}}{\pi \widetilde{\Gamma }}%
\{(g_{f}-g_{c})^{2}+g_{c}(3g_{f}-2g_{c})\frac{-\omega _{c}}{\Omega -\omega
_{c}+4i\gamma } \nonumber \\
&+& g_{c}^{2}\frac{-\omega _{c}(\omega _{f}-\omega _{c})}{%
(\Omega -\omega _{c}+4i\gamma )^{2}}\}
\end{eqnarray}%
The last term carries vanishing spectral weight. The second term on the
r.h.s. represents an impurity induced enhancement ($3g_{f}>2g_{c})$ or
reduction of the bulk conduction electron spin resonance. This contribution
vanishes if $g_{f}=\frac{2}{3}g_{c}$. The static susceptibility takes the
form $\chi _{imp}^{+-}(0)=2\mu _{B}^{2}g_{f}^{2}/(\pi \widetilde{\Gamma })$ .

To summarize, the impurity induced component of the total dynamical spin
susceptibility of a Kondo ion is characterized by a broad excitation peak of
width $\widetilde{\Gamma }\simeq T_{K}$ at $\Omega =\omega _{f}$ and a
narrow peak or dip 
of width $4\gamma $ at $\Omega =\omega _{c}$, where $\gamma $ is the
conduction electron relaxation rate. The relative weights of the two
components depend sensitively on the ratio of $g$-factors. This structure is
not easily detected in an ESR experiment. The narrow resonance line has the
same position and width as the bulk ESR line. Its weight per atom is,
however, enhanced by the large factor $T_{F}/T_{K}$, which comes from the
renormalized susceptibility scale prefactor $\propto 1/\widetilde{\Gamma }$.
Therefore, the ESR response of a diluted magnetic alloy with a concentration
of Kondo ions $c>T_{K}/T_{F}$ will be dominated by the impurity contribution
determined in the above Eqs.\ (6,7).

\section{Anderson lattice model in the Kondo screened regime.}

The Hamiltonian of the simplest Anderson lattice model, assuming momentum
independent hybridization and an isotropic conduction band with flat density
of states is given by 
\begin{eqnarray}
H=H_{c}&+&\sum_{i,\sigma }\epsilon _{f\sigma }f_{i\sigma }^{+}f_{i\sigma
}+U\sum_{i}n_{fi\uparrow }n_{fi\downarrow }\nonumber \\
&+&V\sum_{i,\mathbf{k},\sigma }(e^{i%
\mathbf{{k\cdot R}_{i}}}f_{i\sigma }^{+}c_{\mathbf{k}\sigma }+h.c.)
\end{eqnarray}%
Here $H_{c}$ and $\epsilon _{f\sigma },V,$ $U$ \ have been introduced in
Sec. II and the $\mathbf{{R}_{i}}$ \ are lattice site vectors. The single
particle Green's functions are given by Dyson's equation 
\begin{eqnarray}
&&{\hskip -0.75in}\begin{pmatrix}
i\omega _{n}-\epsilon _{f\sigma }-\Sigma _{f\sigma }(i\omega _{n},\mathbf{k})
& -V \\ 
-V & i\omega _{n}-\epsilon _{\mathbf{k\sigma }}-\Sigma _{c\sigma }(i\omega_n ,%
\mathbf{k})%
\end{pmatrix}%
{\cal G}=1, \nonumber \\
{\rm where} \ \ &{\cal G}&=\begin{pmatrix}
G_{\mathbf{k}\sigma }^{ff} & G_{\mathbf{k}\sigma }^{cf} \\ 
G_{\mathbf{k}\sigma }^{fc} & G_{\mathbf{k}\sigma }^{cc}%
\end{pmatrix}.%
\end{eqnarray}

We assume Fermi liquid theory to hold. Then the self energy $\Sigma
_{f\sigma }(\omega ,\mathbf{k})$\ \ has a power series expansion in $\omega $%
\ at the Fermi surface, and its imaginary part is small \ $\propto \omega
^{2}$, and may be neglected in lowest order. One may use \ $\omega -\epsilon
_{f\sigma }-\Sigma _{f\sigma }(\omega ,\mathbf{k}_{F})=z_{\sigma
}^{-1}[\omega -\widetilde{\epsilon }_{f\sigma }]$, with the quasiparticle
weight factor\ $z_{\sigma }=[1-(\partial \Sigma _{f\sigma }(\omega ,\mathbf{k%
}_{F})/\partial \omega )_{0}]^{-1}$\ and the renormalized energy $\widetilde{%
\epsilon }_{f\sigma }=z_{\sigma }[\epsilon _{d\sigma }+\Sigma _{f\sigma }(0,%
\mathbf{k}_{F})]$. The conduction electron self-energy may be approximated
by $\Sigma _{c\sigma }(\omega +i0,\mathbf{k})=-i\gamma $, where $\gamma $ is
the spin-lattice relaxation rate defined earlier. Then for low energies one
has a quasiparticle description, with $G_{\mathbf{k}\sigma }^{ff}(\omega
)=z_{\sigma }\widetilde{G}_{\mathbf{k}\sigma }^{ff}$, \ $G_{\mathbf{k}\sigma
}^{cf}=\sqrt{z_{\sigma }}\widetilde{G}_{\mathbf{k}\sigma }^{cf}$ and \ the
renormalized \ hybridization amplitude $\widetilde{V}^{2}$\ $=z_{\sigma
}V^{2}$.\ The full matrix of quasiparticle Green's functions is given by 
\begin{equation}
\begin{pmatrix}
\widetilde{G}_{\mathbf{k}\sigma }^{ff} & \widetilde{G}_{\mathbf{k}\sigma
}^{cf} \\ 
\widetilde{G}_{\mathbf{k}\sigma }^{fc} & G_{\mathbf{k}\sigma }^{cc}%
\end{pmatrix}%
=\frac{1}{\mathrm{det}}%
\begin{pmatrix}
\omega -\epsilon _{\mathbf{k\sigma }}+i\gamma  & \widetilde{V} \\ 
\widetilde{V} & \omega -\widetilde{\epsilon }_{f\sigma }%
\end{pmatrix},
\end{equation}%
where $\mathrm{det}=(\omega -\widetilde{\epsilon }_{f\sigma })(\omega
-\epsilon _{\mathbf{k\sigma }}+i\gamma )-\widetilde{V}^{2}=(\omega -\zeta _{%
\mathbf{k\sigma }}^{+})(\omega -\zeta _{\mathbf{k\sigma }}^{-})$. The
complex energy eigenvalues\ are given by 
\begin{eqnarray}
\zeta _{\mathbf{k\sigma }}^{\pm }&=&\frac{1}{2}(\widetilde{\epsilon }_{f\sigma
}+\epsilon _{\mathbf{k\sigma }}-i\gamma )\pm \sqrt{\frac{1}{4}(\widetilde{%
\epsilon }_{f\sigma }-\epsilon _{\mathbf{k\sigma }}+i\gamma )^{2}+\widetilde{%
V}^{2}} \nonumber \\
&=&\epsilon _{\mathbf{k\sigma }}^{\pm }-i\gamma _{\mathbf{k\sigma }%
}^{\pm },
\end{eqnarray}%
where $\epsilon _{\mathbf{k\sigma }}^{\pm }=\mathrm{Re}\ \zeta _{\mathbf{%
k\sigma }}^{\pm }$ and\ $\gamma _{\mathbf{k\sigma }}^{\pm }=-\mathrm{Im}\
\zeta _{\mathbf{k\sigma }}^{\pm }$. We note \ $\gamma _{\mathbf{k\sigma }%
}^{\pm }>0$. There are two energy bands separated by an (indirect) gap ($%
\epsilon _{\mathbf{k\sigma }}^{\min ,\max }$ denote the minimum or maximum
of the conduction band): 
\begin{eqnarray}
\Delta _{g\sigma }&=&\epsilon _{\mathbf{k\sigma }}^{\min }-\epsilon _{\mathbf{%
k\sigma }}^{\max }+\sqrt{(\widetilde{\epsilon }_{f\sigma }-\epsilon _{\mathbf%
{k\sigma }}^{\max })^{2}+4\widetilde{V}^{2}}\nonumber \\
&+&\sqrt{(\widetilde{\epsilon }%
_{f\sigma }-\epsilon _{\mathbf{k\sigma }}^{\min })^{2}+4\widetilde{V}^{2}}%
\gg \omega _{f,c}.
\end{eqnarray}%
We assume that the renormalized $f$-level $\widetilde{\epsilon }_{f\sigma }$ is
inside the conduction band, close to the Fermi level, and consider the case of almost half-filling,
i.e. $n\lesssim 2$ electrons per lattice site. Then only the lower band is
occupied in the ground state. We assume an isotropic band structure for
simplicity. Then near the Fermi level at $k=k_{F}$, the quasiparticle energy
(we drop the spin dependence) has the form 
\begin{eqnarray}
\epsilon _{\mathbf{k}}^{-}-\epsilon _{\mathbf{k}_{F}}^{-}&=&\frac{1}{2}%
(k-k_{F})v_{F}\left[ 1+\frac{(\widetilde{\epsilon }_{f}-\epsilon _{\mathbf{k}%
_{F}})}{\sqrt{(\widetilde{\epsilon }_{f}-\epsilon _{\mathbf{k}_{F}})^{2}+4%
\widetilde{V}^{2}}}\right] \nonumber \\
&\simeq& (k-k_{F})v_{F}^{\ast },
\end{eqnarray}%
where the renormalized Fermi velocity is defined by $v_{F}^{\ast }=v_{F}%
\widetilde{V}^{2}/(\widetilde{\epsilon }_{f}-\epsilon _{\mathbf{k}%
_{F}})^{2}=v_{F}(m/m^{\ast })\simeq v_{F}zV^{2}/(\epsilon _{\mathbf{k}%
_{F}})^{2}$, and we used the fact that $\epsilon _{\mathbf{k}_{F}}\gg 
\widetilde{\epsilon }_{f}$. Note that $\epsilon_{{\bf k}_F}$ is the bare conduction band energy at $k = k_F$, which is far above the true Fermi energy. When $z\ll 1$, the Fermi velocity is
renormalized to very low values and one has a \textquotedblleft heavy
fermion liquid" (effective mass $\ m^{\ast }\gg m$). The effective Fermi
temperature of the heavy quasiparticles is given by $T_{F}^{\ast }=\frac{1}{2%
}k_{F}v_{F\sigma }^{\ast }\ll T_{F}$.

To first order in $\gamma $ the level widths are given by 
\begin{equation}
\gamma _{\mathbf{k\sigma }}^{\pm }=\frac{1}{2}\gamma \left[ 1\mp \frac{%
\widetilde{\epsilon }_{f\sigma }-\epsilon _{\mathbf{k\sigma }}}{\epsilon _{%
\mathbf{k\sigma }}^{+}-\epsilon _{\mathbf{k\sigma }}^{-}}\right] .
\end{equation}

We note that the hybridization induced width $\widetilde{\Gamma }$ of the $f$%
-electron energy level in the impurity problem is now absorbed in the
electronic band structure: the coherent superposition of contributions from
all lattice sites to $\Sigma _{f\sigma }(\omega +i0,\mathbf{k})$ removes the
large constant $i\widetilde{\Gamma }$. The remaining imaginary part of the
self-energy at finite temperatures may be approximated by a constant $\Gamma
_{FL}=cT^{2}/T_{F}^{\ast }$. We shall comment on the effect of quasiparticle
scattering on the ESR-linewidth at the end. Using partial fraction
decomposition, we get the retarded Green function 
\begin{equation}
\widetilde{G}_{\mathbf{k}\sigma }^{ff}(\omega +i0)={\frac{a_{\mathbf{k}%
\sigma }^{ff,+}}{\omega -\zeta _{\mathbf{k\sigma }}^{+}}}+{\frac{a_{\mathbf{k%
}\sigma }^{ff,-}}{\omega -\zeta _{\mathbf{k\sigma }}^{-}}}
\end{equation}%
and similar expressions for $\widetilde{G}_{\mathbf{k}\sigma }^{cf}$ and $G_{%
\mathbf{k}\sigma }^{cc}$, where, with $u_{\mathbf{k}\sigma }=\zeta _{\mathbf{%
k}\sigma }^{+}-\zeta _{\mathbf{k}\sigma }^{-}$, 
\begin{eqnarray*}
a_{\mathbf{k}\sigma }^{ff,\pm }&=&\pm (\zeta _{\mathbf{k\sigma }}^{\pm }-%
\widetilde{\epsilon }_{\mathbf{k}\sigma })/u_{\mathbf{k}\sigma }, \\
 a_{\mathbf{k}\sigma }^{cc,\pm }&=&\pm (\zeta _{\mathbf{k\sigma }}^{\pm }-\epsilon
_{f\mathbf{\sigma }})/u_{\mathbf{k}\sigma }, \\ a_{\mathbf{k}\sigma
}^{cf,\pm }&=&\pm \widetilde{V}/u_{\mathbf{k}\sigma }.
\end{eqnarray*}%
For sufficiently small spin-lattice relaxation, $\gamma \ll (\widetilde{V},%
\widetilde{\epsilon }_{f\sigma })$, we may neglect the imaginary parts in
the weight factors $a_{\mathbf{k}\sigma }^{ff,\pm },...$ and replace $\zeta
_{\mathbf{k\sigma }}^{\pm }$ by $\epsilon _{\mathbf{k\sigma }}^{\pm }$.

The quasiparticles interact via the Fermi liquid interaction. For ESR, the
relevant component of the Fermi liquid interaction is the isotropic
spin-antisymmetric part described by the Landau parameter $F_{0}^{a}$. An
important contribution to $F_{0}^{a}$ comes from the renormalized value $%
\widetilde{U}$ of the bare interaction $U,$ $F_{0}^{a}=-2N_{0}\widetilde{U}$%
, similar to the single impurity case discussed in Sec. II. For the lattice
case, exact numerical results on $\widetilde{U}$ are not available. We note
however, that the onsite repulsion $U$ is likely to be screened down to a
positive value of order $N_{0}^{-1}$, which would lead to a ferromagnetic
Landau parameter $0>F_{0}^{a}\gtrsim-1$. Additional contributions to $%
F_{0}^{a}$ may be generated by nonlocal interactions like the RKKY
interaction, which may be ferro- or antiferromagnetic. We emphasize that this Fermi liquid interaction always leads to a ferromagnetic contribution to the fluctuation spectrum, which may be more or less important depending upon the other contributions.

Following the way in which the interaction was included in the impurity
model, we may express the fully screened $f$-electron susceptibility in
terms of the unscreened one 
\begin{equation}
\chi _{ff}^{+-}(i\Omega _{m})=\chi _{ff,H}^{+-}(i\Omega _{m})/[1-\widetilde{U%
}\chi _{ff,H}^{+-}(i\Omega _{m})],
\end{equation}%
where 
\begin{equation}
\chi _{ff,H}^{+-}(i\Omega _{m})=-T\sum\limits_{\omega _{n}}\sum\limits_{k}%
\widetilde{G}_{\mathbf{k}\downarrow ,H}^{ff}(i\omega _{n}+i\Omega _{m})%
\widetilde{G}_{\mathbf{k}\uparrow ,H}^{ff}(i\omega _{n})
\end{equation}%
The one-to-one correspondence of quasiparticles and bare particles, on which
Landau's Fermi liquid theory rests, allows to calculate the spin
susceptibility from the quasiparticle Green's functions defined above,
without taking the incoherent parts into account. Here the subscript $H$
indicates that the Zeeman energy $\omega _{f}$ is replaced everywhere by 
\begin{equation}
\widetilde{\omega }_{f}=\omega _{f}[1+\widetilde{U}\chi
_{ff}^{+-}(0)]=\omega _{f}[1-\widetilde{U}\chi _{ff,H}^{+-}(0)]^{-1}
\end{equation}

Using the representation of $\widetilde{G}_{\mathbf{k}\downarrow ,H}^{ff}$
in terms of eigenstates, the summation in Eq.\ (17) on \ $\omega _{n}$ and $%
\mathbf{k}$ may be done. In the case that only the lower band is occupied,
the low frequency response is given by, see the Appendix, Eq.\ (A4):
\begin{equation}
\chi ^{+-}(\Omega + i0)={\frac{\chi^{+-}(0)(-\omega _{r}+i\gamma _{r})}{\Omega -\omega _{r}+i\gamma _{r}}},
\end{equation}%
where $\chi^{+-}(0)$ is defined in the Appendix, Eq.\ (A5). The mean field shift largely
cancels out of the resonance frequency
\begin{equation}
\omega _{r}=\frac{1}{2}\omega _{f}\left[ 1-{\frac{(\widetilde{\epsilon }%
_{f}-\epsilon _{\mathbf{k}_{F}})}{\sqrt{(\widetilde{\epsilon }_{f}-\epsilon
_{\mathbf{k}_{F}})^{2}+4\widetilde{V}^{2}}}}\right] \simeq \omega
_{f}(1-m/m^{\ast }),
\end{equation}%
The linewidth, however, is reduced by the exchange interaction, provided the
interaction is ferromagnetic.
\begin{eqnarray}
\gamma_{r}&=&\gamma\left [1+\frac{\widetilde{\epsilon}%
_{f}-\epsilon_{\mathbf{k}_{F}}}{\sqrt{(\widetilde{\epsilon}_{f}-\epsilon _{%
\mathbf{k}_{F}})^{2}+4\widetilde{V}^{2}}}\right ][1-\widetilde{U}%
\chi_{ff,H}^{+-}(0)] \nonumber \\
&\simeq& 2\gamma\frac{m}{m^{\ast}}[1-\widetilde{U}%
\chi_{ff,H}^{+-}(0)]\ll \gamma.
\end{eqnarray}
It is seen that the main narrowing
mechanism is provided by the hybridization through the renormalized
amplitude $\widetilde{V}$, which gives the small factor $m/m^*$. In simple terms, the quasiparticles at the Fermi surface have mainly $f$-character, with only a small admixture (fraction $m/m*$) of conduction electron component. Since only the conduction electrons feel the spin lattice relaxation, the total spin relaxation is a fraction  $m/m*$ of the spin lattice relaxation. Vertex corrections to the spin-lattice relaxation are likely to increase $\gamma_r$ somewhat as they do in the impurity case, Appendix Eq.\ (A1), where $2\gamma$ becomes $4\gamma$.

In order to discuss the temperature and magnetic field dependence of the
linewidth it is necessary to incorporate quasiparticle scattering effects \cite {overhauser}  and inelastic contributions to the spin-lattice relaxation.
In the case that the $g$-factors are sufficiently different, the
contribution to the linewidth from quasiparticle scattering will vary with temperature as $
T^{2}/T_{F}^{\ast}$ and with magnetic field $H$ as $H^{2}/T_{F}^{\ast}$. In
the case of equal $g$-factors the latter contribution will be cancelled by
vertex corrections. Additional temperature dependence may arise from coupling to phonons.

\section{Conclusion}
This paper is motivated by the recent observations of electron spin resonance at low temperature in some heavy-fermion compounds. We have calculated the dynamical susceptibility, which describes the resonance, at low temperature in the fully screened Kondo regime for both a single Kondo impurity spin as well as for the Kondo lattice, described here by the Anderson lattice model.

We have not addressed the behavior of the susceptibilities at temperatures in the neighborhood of the Kondo temperature, where linewidths are expected to be very large due to rapid spin fluctuations in that temperature range. Rather, we deal with the very low temperature regime, where a Kondo impurity is fully screened and where the heavy-electron Fermi liquid has formed in the Anderson lattice.

For the realistic case in which the $g$-factors of $f$ electrons and conduction electrons are different, we find for the single impurity that structure persists at both the $f$ electron and conduction electron resonance frequencies. The impurity resonance continues to have a large width, of order $T_K$, while for the conduction electron resonance there is an impurity-induced contribution that increases or decreases the amplitude depending on the ratio of $g$-factors.

The situation is quite different for the lattice case. Here, the hybridization of the $f$ and conduction electrons and Fermi-liquid interaction lead to modifications of the susceptibility that can lead to substantial line narrowing and hence the possibility of experimental observation. We find a sharp ESR line near the underlying local $f$ electron resonance. The line is substantially narrowed by a factor of the mass ratio $m/m*$ and by the effect of the Fermi liquid interaction $F^0_a$ provided it is negative (ferromagnetic).

We note that the ESR has been only been seen in heavy fermion compounds for which there is independent evidence for ferromagnetic fluctuations \cite{Ce,YRS2}. We
suggest that our analysis accounts for this observation.

\begin{acknowledgements}
We thank the Aspen Center for Physics, where this work was begun and completed. We thank Q. Si for pointing us to this problem and we acknowledge helpful discussions with K. Baberschke, P. Coleman, I. Martin, D. Pines. We are grateful to J. Sichelschmidt for his generous sharing of information and data. This work has been supported in part by the DFG-Center for Functional Nanostructurres and the DFG-Forschergruppe "Quantum Phase Transitions" (PW).
\end{acknowledgements}

\appendix*
\section{}

\subsection{Anderson impurity model in the Kondo screened regime: Green's
function approach to $\chi^{+-}(\Omega)$}

As derived in the main text, Eq.\ (2), the Green's functions of conduction electrons and
local electrons, including the Fermi-liquid interaction, are given by
\begin{eqnarray*}
G_{f\sigma}(i\omega_{n})&=&[i\omega_{n}+\sigma\omega_{f}+i\widetilde\Gamma
sign(\omega_{n})]^{-1}, \\
G_{c\sigma}({\bf k},i\omega_{n})&=&[i\omega_{n}-\varepsilon_{k}+\sigma
\omega_{c}/2+i\gamma sign(\omega_{n})]^{-1}.
\end{eqnarray*}

The dynamical transverse susceptibility $\chi^{+-}(\Omega)$, where $\Omega$
is the frequency of an a.c. electromagnetic field polarized transverse to the
static magnetic field, is given by
\[
\chi^{+-}(\Omega)=\mu_{B}^{2}[g_{c}^{2}\chi_{cc}^{+-}(\Omega)+g_{f}^{2}%
\chi_{ff}^{+-}(\Omega)+2g_{c}g_{f}\chi_{cf}^{+-}(\Omega)].
\]

The
partial susceptibilities are obtained by evaluating Feynman bubble diagrams dressed by vertex
corrections of the ladder type referring to the Fermi liquid interaction
(local electrons) and the spin-orbit interaction (impurity correlation lines
for the conduction electrons).

The local susceptibility in the absence of vertex corrections is obtained as
\begin{eqnarray*}
\chi_{ff,H}^{+-}(i\Omega_{m})&=&-T
\sum_{\omega_{n}} G_{f\downarrow}(i\omega_{n}+i\Omega_{m})G_{f\uparrow}(i\omega_{n}) \\
&=&\frac
{2}{\pi \widetilde \Gamma}\frac{-\omega_{f}+i \widetilde\Gamma}{i\Omega_{m}-2\omega_{f}+2i \widetilde\Gamma}.
\end{eqnarray*}
The vertex
corrections are obtained from the Bethe-Salpeter equation
\[
\Lambda(i\Omega_{m})=1+\widetilde{U}\chi_{ff,H}^{+-}(i\Omega_{m}%
)\Lambda(i\Omega_{m})=\frac{\Omega-2\omega_{f}+2i \widetilde\Gamma}{\Omega-\omega
_{f}+i \widetilde\Gamma},
\]
where
we used $\widetilde{U}=\pi\widetilde\Gamma$. Then
\[
\chi_{ff}^{+-}(i\Omega_{m})=\chi_{ff,H}^{+-}(i\Omega_{m})\Lambda(i\Omega
_{m})=\frac{2}{\pi \widetilde\Gamma}\frac{-\omega_{f}+i \widetilde\Gamma}{i\Omega_{m}-\omega
_{f}+i \widetilde\Gamma}.
\]

The conduction electron susceptibility consists of four contributions
\[
\chi_{cc}^{+-}(i\Omega_{m})=\chi_{cc}^{bulk}(i\Omega_{m})+\sum_{i=1}^3\chi_{cc}%
^{(i)}(i\Omega_{m})
\]
 The
bulk contribution has the form
$\chi_{cc}^{bulk}(i\Omega_{m})=N\chi_{cc}^{0}(\Omega+i0) \Phi(i\omega
_{n},i\Omega_{m})$,
where
$N$ is the number of atoms in the system and $ \chi_{cc}^{0}%
(\Omega+i0)=-T%
\sum_{\omega_{n},{\bf k}}
G_{c\downarrow}({\bf k},i\omega_{n}+i\Omega_{m})G_{c\uparrow}({\bf k},i\omega_{n}%
)=N_{0}(-\omega_{c}+2i\gamma)/(\Omega-\omega_{c}+2i\gamma)$,
where\ $N_{0}$ is the
conduction electron density of states at the Fermi level. The vertex function
$\Phi(i\omega_{n},i\Omega_{m})$ is found as solution to the equation:
\begin{widetext}
\begin{equation*}
\Phi(i\omega_{n},i\Omega_{m})=1-h^{2}%
\sum_{{\bf k}}
G_{c\downarrow}^{0}({\bf k},i\omega_{n}+i\Omega_{m})G_{c\uparrow}^{0}({\bf k},i\omega
_{n})\Phi(i\omega_{n},i\Omega_{m})
\end{equation*}
as 
\[
\Phi(i\omega_{n},i\Omega_{m})=\theta(-\omega_{n})\theta(\omega_{n}+\Omega
_{m})\frac{i\Omega_{m}-\omega_{c}+2i\gamma}{i\Omega_{m}-\omega_{c}+4i\gamma
}+[1-\theta(-\omega_{n})\theta(\omega_{n}+\Omega_{m})]
\]
\end{widetext}
 Note that the minus sign in
front of $h^{2}$ is generated by the Pauli matrices that appear in $H_c$, Eq.\ (3) of the main text (in the case of potential
scattering there would be no sign change): $\sum_{i,\alpha,\beta}
\sigma_{\alpha\downarrow}^{i}\sigma_{\uparrow\beta}^{i}G_{c\alpha}%
^{0}G_{c\beta}^{0}=-G_{c\downarrow}^{0}G_{c\uparrow}^{0}$ . As a
consequence, the vertex corrections double the linewidth: $2\gamma
\rightarrow4\gamma$ . In the case of potential scattering the vertex
corrections cancel the self-energy induced linewidth, so that potential
scattering does not contribute to the spin relaxation, as expected. Combining
the above results we find
\bq
\chi_{cc}^{bulk}(\Omega+i0)=NN_{0}\frac{-\omega_{c}+4i\gamma
}{\Omega-\omega_{c}+4i\gamma}.
\eq
 The
remaining contributions are obtained from 
\begin{widetext}\begin{eqnarray*}
\chi_{cc}^{(1)}(i\Omega_{m})
&=& -V^{2}T\sum_{\omega_n}\sum_{\bf k}\Big\{ [G_{c\downarrow}^{0}({\bf k},i\omega_{n}+i\Omega_{m})]^{2}G_{c\uparrow}^{0}({\bf k},i\omega_n)G_{f\downarrow}(i\omega_n+i\Omega_m)  \\
&+& G_{c\downarrow}^{0}({\bf k},i\omega_n
+i\Omega_m)[G_{c\uparrow}^{0}({\bf k},i\omega_n)]^2 G_{f\uparrow}(i\omega_n) \Big\}
,
\\
\chi_{cc}^{(2)}(i\Omega_{m}) &=& -T\sum_{\omega_n}
\left[  V^{2}
\sum_{\bf k}
G_{c\downarrow}^{0}({\bf k},i\omega_n + i\Omega_m)G_{c\uparrow}^{0}({\bf k},i\omega
_n)\Phi(i\omega_n,i\Omega_m)\right]^2 G_{f\downarrow}(i\omega
_n+i\Omega_m)G_{f\uparrow}(i\omega_n),   \\
\chi_{cc}^{(3)}(i\Omega_m) &=& -\left[  V^{2}T\sum_{\omega_n}
\sum_{\bf k} G_{c\downarrow}^{0}({\bf k},i\omega_n + i\Omega_m)G_{c\uparrow}^0({\bf k},i\omega_n)\Phi(i\omega_n,i\Omega_m)G_{f\downarrow}(i\omega_n+i\Omega
_m)G_{f\uparrow}(i\omega_n)\right ] ^2  \\
&\times& [- \widetilde{U}\Lambda(i\Omega_m)]. 
\end{eqnarray*}
Using
\[
\sum_{\bf k}
[G_{c\downarrow}^{0}({\bf k},i\omega_n+i\Omega_m)]^2 G_{c\uparrow}
^0({\bf k},i\omega_n) = N_0\frac{2\pi i}{(i\Omega_{m}-\omega_{c}+2i\gamma)^{2}%
} = -\sum_{\bf k} G_{c\downarrow}^0 ({\bf k},i\omega_n+i\Omega_m)[G_{c\uparrow}^0({\bf k},i\omega_n)]^{2}
\]
and the identity
\[
G_{f\downarrow}(i\omega_{n}+i\Omega_{m})-G_{f\uparrow}(i\omega_{n}%
)=-(i\Omega_{m}-2\omega_{f}+2i \widetilde\Gamma)G_{f\downarrow}(i\omega_{n}+i\Omega
_{m})G_{f\uparrow}(i\omega_{n})
\]
as well as
\[
\Pi(i\Omega_{m})=T\!\!\!\!\!\!\sum_{-\Omega_{m}<\omega_{n}<0}
G_{f\downarrow}(i\omega_{n}+i\Omega_{m})G_{f\uparrow}(i\omega_{n})=\frac
{1}{\pi \widetilde\Gamma}\frac{i\Omega_{m}}{i\Omega_{m}-2\omega_{f}+2i \widetilde\Gamma}
\]
we get
\begin{eqnarray*}
\chi_{cc}^{(1)}(i\Omega_{m}) &=& \frac{2}{\pi \widetilde\Gamma}\frac{\Omega}{\Omega
-2\omega_{f}+2i \widetilde\Gamma} \\
\chi_{cc}^{(2)}(i\Omega_{m})&=&\frac{2}{\pi\widetilde\Gamma}\frac{2\widetilde\Gamma^{2}
}{(i\Omega_{m}-\omega_{c}+4i\gamma)^{2}}\frac{\Omega}{\Omega-2\omega
_{f}+2i\widetilde\Gamma} \\
\chi_{cc}^{(3)}(i\Omega_{m})& =& -\frac{2}{\pi\widetilde\Gamma}\frac{\Omega^{2}}%
{(i\Omega_{m}-\omega_{c}+4i\gamma)^{2}}\frac{\widetilde\Gamma^{2}}{(\Omega-2\omega
_{f}+2i\widetilde\Gamma)(\Omega-2\omega_{f}+2i\widetilde\Gamma)}
\end{eqnarray*}
Adding
the three contributions we find
\bq
\sum_{i=1}^{3}
\chi_{cc}^{(i)}(i\Omega_{m})=\frac{2}{\pi\widetilde\Gamma}\frac{\Omega(\Omega-\omega
_{f})}{(i\Omega_{m}-\omega_{c}+4i\gamma)^{2}}\frac{i\widetilde\Gamma}{\Omega-\omega
_{f}+i\widetilde\Gamma}
\eq

The mixed susceptibility may be calculated from the bubble diagram beginning
with a conduction electron particle-hole line and ending with a local electron
p-h line, dressed by vertex corrections at both ends:
\begin{eqnarray}
\chi_{cf}^{+-}(\Omega_{m}) &=& -T\sum_{\omega_{n},{\bf k}}
G_{c\downarrow}^{0}({\bf k},i\omega_{n}+i\Omega_{m})G_{c\uparrow}^{0}({\bf k},i\omega
_{n})\Phi(i\omega_{n},i\Omega_{m})V^{2}G_{f\downarrow}(i\omega_{n}+i\Omega
_{m})G_{f\uparrow}(i\omega_{n})\Lambda(i\Omega_{m}) \nonumber \\
&=& \frac{2}{\pi\widetilde\Gamma}\frac{-i\widetilde\Gamma}{i\Omega
_{m}-\omega_{c}+4i\gamma}\frac{i\Omega_{m}}{i\Omega_{m}-2\omega_{f}+2i\widetilde\Gamma}\end{eqnarray}
\end{widetext}
After analytical continuation to the real frequency axis and combining the
contributions the total impurity susceptibility is obtained as given in the
main text, Eqs. (6,7).

\subsection{Anderson lattice model in the Kondo screened regime: Green's function
approach to $\chi^{+-}(\Omega)$.}

As derived in the main text Eq.\ (10), the matrix of quasiparticle Green's functions is
given by
\[
\begin{pmatrix}
\widetilde{G}_{{\bf k}\sigma }^{ff} & \widetilde{G}_{{\bf k}\sigma
}^{cf} \\ 
\widetilde{G}_{{\bf k}\sigma }^{fc} & G_{{\bf k}\sigma }^{cc}%
\end{pmatrix}%
=\frac{1}{{\rm det} }
\begin{pmatrix}
\omega -\epsilon _{\bf{k\sigma }}+i\gamma  & \widetilde{V} \\ 
\widetilde{V} & \omega -\widetilde{\epsilon }_{f\sigma }%
\end{pmatrix},%
\]
where ${\rm det} =(\omega -\widetilde{\epsilon }%
_{f\sigma })(\omega -\epsilon _{\bf{k\sigma }}+i\gamma )-\widetilde{V}%
^{2}=(\omega -\zeta _{\bf{k\sigma }}^{+})(\omega -\zeta _{\bf{%
k\sigma }}^{-})$.  The complex energy eigenvalues\ are given by
\begin{eqnarray*}
\zeta _{\bf{k\sigma }}^{\pm }&=&\frac{1}{2}(\widetilde{\epsilon }%
_{f\sigma }+\epsilon _{\bf{k\sigma }}-i\gamma )\pm \sqrt{\frac{1}{4}(%
\widetilde{\epsilon }_{f\sigma }-\epsilon _{\bf{k\sigma }}+i\gamma )^{2}+%
\widetilde{V}^{2}} \\
&=&\epsilon _{\bf{k\sigma }}^{\pm }-i\gamma _{\bf{%
k\sigma }}^{\pm },
\end{eqnarray*}
 where, expanding to
leading order in $\gamma$, as well as in $\omega_{f},\omega_{c}$,
\begin{eqnarray*}
\epsilon_{\mathbf{k\sigma}}^{\pm}=\operatorname{Re}\zeta_{\mathbf{k\sigma}%
}^{\pm}&\simeq&\frac{1}{2}(\widetilde{\epsilon}_{f\sigma}+\epsilon
_{\mathbf{k\sigma}})\pm\sqrt{\frac{1}{4}(\widetilde{\epsilon}_{f\sigma
}-\epsilon_{\mathbf{k\sigma}})^{2}+\widetilde{V}^{2}} \\
&\simeq&\epsilon
_{\mathbf{k}}^{\pm}-\frac{1}{2}\omega_{\mathbf{k}}^{\pm}\sigma
\end{eqnarray*}
and
\[
\gamma_{\mathbf{k\sigma}}^{\pm}=-\operatorname{Im}\zeta
_{\mathbf{k\sigma}}^{\pm}\simeq\frac{1}{2}\gamma\lbrack1\pm\frac
{\widetilde{\epsilon}_{f\sigma}-\epsilon_{\mathbf{k\sigma}}}{\epsilon
_{\mathbf{k\sigma}}^{+}-\epsilon_{\mathbf{k\sigma}}^{-}}]\simeq\gamma
_{\mathbf{k}}^{\pm}-\frac{1}{2}\eta_{\mathbf{k}}^{\pm}\sigma,
\]
with
\begin{eqnarray*}
\epsilon_{\mathbf{k}}^{\pm}&=&\frac{1}{2}(\widetilde{\epsilon}_{f}%
+\epsilon_{\mathbf{k}})\pm\frac{1}{2}\sqrt{(\widetilde{\epsilon}_{f}%
-\epsilon_{\mathbf{k}})^{2}+4\widetilde{V}^{2}} \\
\omega_{\mathbf{k}}^{\pm}&=&\frac{1}{2}(\widetilde{\omega}_{f}+\omega_{c}%
)\pm\frac{1}{2}\frac{\widetilde{\epsilon}_{f}-\epsilon_{\mathbf{k}}}%
{\sqrt{(\widetilde{\epsilon}_{f}-\epsilon_{\mathbf{k}})^{2}+4\widetilde{V}%
^{2}}}(\widetilde{\omega}_{f}-\omega_{c}) \\
\gamma_{\mathbf{k}}^{\pm}&=&\frac{1}{2}\gamma\lbrack1\mp\frac
{\widetilde{\epsilon}_{f}-\epsilon_{\mathbf{k}}}{\sqrt{(\widetilde{\epsilon
}_{f}-\epsilon_{\mathbf{k}})^{2}+4\widetilde{V}^{2}}}] \\
 \eta
_{\mathbf{k}}^{\pm}&=&\mp\frac{1}{2}\gamma\frac{\widetilde{\omega}_{f}%
-\omega_{c}}{\sqrt{(\widetilde{\epsilon}_{f}-\epsilon_{\mathbf{k}}%
)^{2}+4\widetilde{V}^{2}}}\frac{4\widetilde{V}^{2}}{(\widetilde{\epsilon}%
_{f}-\epsilon_{\mathbf{k}})^{2}+4\widetilde{V}^{2}}.
\end{eqnarray*}

The susceptibility is, as in the case of the impurity, given by the sum of three
contributions: $ff$, $cc$ and ($cf,fc$):
\[
\chi^{+-}(\Omega)=\mu_{B}^{2}[g_{c}^{2}\chi_{cc}^{+-}(\Omega
)+g_{f}^{2}\chi_{ff}^{+-}(\Omega)+2g_{c}g_{f}\chi_{cf}^{+-}(\Omega)].
\]
 Here
the $ff$-susceptibility is screened by the Fermi liquid interaction
\[
\chi_{ff}^{+-}(i\Omega_{m})=\chi_{ff,H}^{+-}(i\Omega_{m})\Lambda(i\Omega
_{m}), 
\]
 with
  \[
 \Lambda(i\Omega_{m})=1/[1-\widetilde{U}\chi_{ff,H}%
^{+-}(i\Omega_{m})]
\]
 where
\[
\chi_{ff,H}^{+-}(i\Omega_{m})=-T{\textstyle\sum\limits_{\omega_{n},{\bf k}}}
\widetilde{G}_{\mathbf{k}\downarrow,H}^{ff}(i\omega_{n}+i\Omega_{m}%
)\widetilde{G}_{\mathbf{k}\uparrow,H}^{ff}(i\omega_{n})
\]
 Similarly,
\[
\chi_{cf}^{+-}(i\Omega_{m})=\chi_{cf,H}^{+-}(i\Omega_{m})\Lambda(i\Omega
_{m}),
\]
 where
\[
 chi_{cf,H}^{+-}(i\Omega_{m})=-T{\textstyle\sum\limits_{\omega_{n},{\bf k}}}
\widetilde{G}_{\mathbf{k}\downarrow}^{cf}(i\omega_{n}+i\Omega_{m}%
)\widetilde{G}_{\mathbf{k}\uparrow}^{cf}(i\omega_{n})
\]
and
\[
\chi_{cc}^{+-}(i\Omega_{m})=\chi_{cc,H}^{+-}(i\Omega_{m})+\widetilde{U}%
[\chi_{cd,H}^{+-}(i\Omega_{m})]^{2}\Lambda(i\Omega_{m}), 
\]
where
\[ 
\chi_{cc,H}^{+-}(i\Omega_{m})=-T{\textstyle\sum\limits_{\omega_{n},{\bf k}}}
G_{\mathbf{k}\downarrow}^{cc}(i\omega_{n}+i\Omega_{m})G_{\mathbf{k}\uparrow
}^{cc}(i\omega_{n})
\]
Using
the representation of the Green's functions in terms of the eigenstates
$\nu=\pm$ , and the fact that low energy excitations are only possible close
to the Fermi energy, which we assume to lie in the lower band ($\nu=-$),
only the ($-$)-components contribute to $\chi_{ij,H}^{+-}(i\Omega_{m}):$
\[
\chi_{ff,H}^{+-}(\Omega+i0)=-%
{\textstyle\sum\limits_{\bf k}}
a_{\mathbf{k}\downarrow}^{ff,-}a_{\mathbf{k}\uparrow}^{ff,-}\frac
{f(\zeta_{\mathbf{k\uparrow}}^{-})-f(\zeta_{\mathbf{k\downarrow}}^{-})}%
{\Omega-\zeta_{\mathbf{k\downarrow}}^{-}+\zeta_{\mathbf{k\uparrow}}^{-}+i0},
\]
where
in the arguments of the Fermi function $f(\epsilon)$, the complex-valued
energy\ \ $\zeta_{\mathbf{k\downarrow}}^{-}$ appears. Employing $\zeta_{\mathbf{k\downarrow}}^{-}-\zeta_{\mathbf{k\uparrow}%
}^{-}\simeq\omega_{\mathbf{k}}^{-}-2i\gamma_{\mathbf{k}}^{-}$, and $%
{\textstyle\sum\limits_{k}}
[f(\zeta_{\mathbf{k\uparrow}}^{-})-f(\zeta_{\mathbf{k\downarrow}}%
^{-})]\simeq%
{\textstyle\sum\limits_{k}}
(\partial f/\partial\epsilon_{k}^{-})(-\omega_{\mathbf{k}_{F}}^{-}%
+2i\gamma_{\mathbf{k}_{F}}^{-})=N_{0}(\omega_{\mathbf{k}_{F}}^{-}%
-2i\gamma_{\mathbf{k}_{F}}^{-})$, we get
\[
\chi_{ff,H}^{+-}(\Omega+i0)=N_{0}a_{\mathbf{k}_{F}\downarrow
}^{ff,-}a_{\mathbf{k}_{F}\uparrow}^{ff,-}\frac{-\omega_{\mathbf{k}_{F}}%
^{-}+2i\gamma_{\mathbf{k}_{F}}^{-}}{\Omega-\omega_{\mathbf{k}_{F}}%
^{-}+2i\gamma_{\mathbf{k}_{F}}^{-}} 
\]
and hence
\[
\chi_{ff,H}%
^{+-}(0)=N_{0}a_{\mathbf{k}_{F}\downarrow}^{ff,-}a_{\mathbf{k}%
_{F}\uparrow}^{ff,-}.
\] 
 Equivalent
expressions hold for the $ff$ and $cf$ components. The vertex function follows as
\[
\Lambda(\Omega+i0)=\frac{\Omega-\omega_{\mathbf{k}_{F}}^{-}+2i\gamma
_{\mathbf{k}_{F}}^{-}}{\Omega-(\omega_{\mathbf{k}_{F}}^{-}-2i\gamma
_{\mathbf{k}_{F}}^{-})(1-\widetilde{U}\chi_{ff,H}^{+-}(0))}
\]
and
the renormalized $ff$-susceptibility takes the form
\begin{eqnarray*}
\chi_{ff}^{+-}(\Omega+i0)&=&\chi_{ff}^{+-}(0)\frac{-\omega_{r}+i\gamma_{r}%
}{\Omega-\omega_{r}+i\gamma_{r}},  \ {\rm where} \\
\omega_{r}-i\gamma_{r}&=&(\omega_{\mathbf{k}_{F}}^{-}-2i\gamma_{\mathbf{k}_{F}%
}^{-})(1-\widetilde{U}\chi_{ff,H}^{+-}(0)) \  {\rm and} \\
\chi_{ff}^{+-}(0)&=&\chi_{ff,H}^{+-}(0)/[1-\widetilde{U}\chi_{ff,H}^{+-}(0)]
\end{eqnarray*}
as
discussed in the main text, Eqs.\ (20,21). The total susceptibility consists of two resonance terms:
\begin{widetext}
\begin{equation}
\chi^{+-}(\Omega+i0) = \chi_{r}^{+-}(0)\frac{-\omega_{r}%
+i\gamma_{r}}{\Omega-\omega_{r}+i\gamma_{r}} + g_{c}^{2}\frac{-\omega
_{\mathbf{k}_{F}}^{-}+2i\gamma_{\mathbf{k}_{F}}^{-}}{\Omega-\omega
_{\mathbf{k}_{F}}^{-}+2i\gamma_{\mathbf{k}_{F}}^{-}}\{\chi_{cc,H}%
^{+-}(0)+\widetilde{U}[\chi_{cf,H}^{+-}(0)]^{2}\frac{-\omega_{r}+i\gamma_{r}%
}{\Omega-\omega_{r}+i\gamma_{r}}\},
\end{equation}
where $\chi_{r}^{+-}(0)=g_{f}^{2}\chi_{ff}^{+-}(0)+2g_{c}g_{f}\chi_{cf}^{+-}(0)$. In
the case that $\gamma_{{\bf k}_F}^- \gg \gamma_r$, the resonance part simplifies to
\bq
\chi^{+-}(\Omega+i0)=\chi^{+-}(0)\frac{-\omega_{r}+i\gamma_{r}%
}{\Omega-\omega_{r}+i\gamma_{r}}, \ \ {\rm where} \ \ \chi^{+-}(0) = \chi_{r}^{+-}(0) + g_c^2\widetilde U[\chi_{cf,H}^{+-}(0)]^{2}
\eq
\end{widetext}

\end{document}